\documentclass[12pt]{article}
\usepackage{amssymb}
\begin{document}
\begin{flushright}
\baselineskip=12pt
{hep-th/0406031}\\
{June  2004}
\end{flushright}

\def\beq{\begin{equation}}
\def\bea{\begin{eqnarray}}
\def\eeq{\end{equation}}
\def\eea{\end{eqnarray}}
\def\IR{I\kern-.4em R}
\def\npb{\it Nuclear Physics B}
\def\th{arXiv:hep-th/}
\def\ph{arXiv:hep-ph/}

\begin{center}
\vglue 1cm
{\LARGE \bf Non-minimal Higgs content in  standard-like models from D-branes at a $\mathbb{Z} _N$  singularity \\} 
\vglue 1cm
{D. BAILIN$^{\clubsuit}$\footnote
{D.Bailin@sussex.ac.uk}
  and A. LOVE$^{\diamondsuit}$ \\}
\vglue 0.2cm
	{$\clubsuit$ \it  Department of Physics \& Astronomy, University of Sussex\\}
{\it Brighton BN1 9QH, U.K. \\}

\vglue 0.2cm
{$\diamondsuit$ \it  Centre for Particle Physics, Royal Holloway, University of London \\}
{\it Egham,  Surrey TW20 0EX, U.K. }
\baselineskip=12pt
\vglue 2.5cm

 \vglue 2.5cm
ABSTRACT
\end{center}
We show that attempts to construct the standard model, or the MSSM, by placing D3-branes and D7-branes
 at a $\mathbb{Z} _N$ orbifold or 
orientifold singularity all require that the  electroweak Higgs content  is non-minimal. 
For the orbifold the lower bound on the number $n(H) + n({\bar{H}})$ of electroweak Higgs doublets is the number 
$n(q^c_L)=6$ of quark singlets, and for the orientifold the lower bound can be one less. 
As a consequence there is a generic flavour changing neutral current problem in such models.

\vglue 0.5cm
{\rightskip=3pc
\leftskip=3pc
\noindent
\baselineskip=20pt

}

\vfill\eject
\setcounter{page}{1}
\pagestyle{plain}
\baselineskip=14pt
The construction of models that  lead to something like the standard model has been a major activity in string theory for many 
years. (See, for example, \cite{Bailin:2001ia} for a review.) 
The D-brane world offers an attractive, bottom-up route to getting standard-like models from Type II string theory  that 
has been particularly popular  recently. 
Open strings that begin and end on a stack of $M$ D-branes generate
 the gauge bosons of the group $U(M)$
 living in the world volume of the D-branes.
  Recently ``intersecting brane'' models have enjoyed considerable popularity. 
  (See \cite{Lust:2004ks} for a recent review.)
 In these models  one
 starts with one stack of 3 D-branes, another of 2, and $n$ other stacks each having just 1 D-brane,
 thereby generating 
the gauge group $U(3) \otimes U(2) \otimes U(1)^n$. The D4-,5- or 6-branes wrap the three large spatial dimensions
 and respectively  1-, 2- or 3-cycles of the six-dimensional internal space (typically a torus $T^6$ or a Calabi-Yau 3-fold)
   on which the theory is compactified.
Then fermions in bi-fundamental
 representations of the corresponding gauge groups can arise at the multiple intersections of such stacks \cite{Berkooz:1996km},
  but to get $D=4$ {\it chiral} fermions 
  the intersecting branes should sit at a singular point in the space transverse to the branes, an orbifold fixed point, for example. 
  In general,
   such configurations yield a non-supersymmetric spectrum, so that, to avoid the hierarchy problem, the
    string scale associated with such models must be no more than a few TeV. If so, some of 
    the compactified dimensions must be large, and this raises the question of how  
     the required high Planck energy scale associated with gravitation emerges.
      Provided that the volume of the space orthogonal to the 
     wrapped space is sufficiently large,  it seems that both scales can be accommodated \cite{Blumenhagen:2002wn, Uranga:2002pg}.  
    However, a generic feature of these  models is that flavour changing neutral currents are
     generated  by four-fermion operators induced by string instantons \cite{Abel:2003yh}. 
      Although such  operators 
      allow the emergence of a realistic  pattern  of fermion masses and mixing angles, 
      the severe experimental limits on  flavour changing neutral currents require that  the string 
      scale is rather high, of order 10$^4$ TeV. This makes the fine tuning problem very severe 
      and renders the viability of these models highly questionable. 
    In a non-supersymmetric theory the cancellation of the Ramond-Ramond (RR) tadpoles does 
{\it not} ensure the cancellation of the Neveu-Schwarz-Neveu-Schwarz (NSNS) tadpoles. NSNS tadpoles are 
simply the first derivatives of the scalar potential with repect to the scalar fields, specifically  the complex 
structure moduli and the dilaton. Thus a consequence of the non-cancellation is that 
 there is an instability in the complex structure moduli \cite{Blumenhagen:2001te}.
 One  way to stabilise  the complex structure moduli is to use an orbifold, rather than a torus, 
for the space wrapped by the D-branes. If the embedding is supersymmetric, RR tadpole 
cancellation ensures the cancellation of the NSNS tadpoles too \cite{Cvetic:2001tj, Cvetic:2001nr}.
Using D6-branes and a $\mathbb{Z} _4, \mathbb{Z}_4 \times \mathbb{Z}_2$ or $\mathbb{Z}_6$  orientifold, some or all of the 
 complex structure moduli may be stabilised \cite{Blumenhagen:2002gw,Honecker:2003vq,Honecker:2004kb}. 
Although  a semi-realistic  three-generation model has been obtained  this way \cite{Honecker:2004kb}, it has non-minimal Higgs 
content, so it too will have flavour changing neutral currents.

 For all of these reasons it seems timely to re-examine the viability of the earlier ``bottom-up'' models  
 \cite{Aldazabal:2000sa, Bailin:2000kd, Berenstein:2001nk, Alday:2002uc}, 
 the study of which began before the intersecting brane models became popular.   
 In these  one starts with a set of $n+5$ D3-branes  situated at  an orbifold $T ^6/ \mathbb{Z} _N$ singularity. 
 In contrast to the intersecting brane models discussed above, 
 the  gauge group $U(3) \otimes U(2) \otimes U(1)^n$ is obtained by choosing a suitable embedding $\gamma _{\theta,3}$ of the 
 action of the  generator $\theta$ of the point group $\mathbb{Z} _N$ on the Chan-Paton indices of the D3-branes. 
 Before the orbifold projection, a set of $n+5$ D3-branes  generically gives an ${\mathcal N}=4$ supersymmetric $U(n+5)$ gauge theory. 
In ${\mathcal N}=1$ language this consists of an adjoint vector multiplet and 3 adjoint chiral multiplets. 
In terms of component fields there are $U(n+5)$ gauge fields, four adjoint fermions, transforming as the ${\bf 4}$ 
representaion of  $SU(4)$, and six 
adjoint scalars,  transforming as the ${\bf 6}$ of $SU(4)$.
The action of $\theta$ on the four fermions is given by
\beq
R_4(\theta)= {\rm diag}\left( e^{2\pi i a_1/N}, e^{2\pi i a_2/N}, e^{2\pi i a_3/N}, e^{2\pi i a_4/N}\right) \label{thetaferm}
\eeq
where the $a_{\alpha}$ are integers satisfying
\beq 
a_1+a_2+a_3+a_4=0 \bmod N \label{thetaferm2}
\eeq
The action of the  point group  on the six scalars is given by
\beq
R_6(\theta)= {\rm diag}\left( e^{2\pi i b_1/N}, e^{-2\pi i b_1/N},e^{2\pi i b_2/N},e^{-2\pi i b_2/N}, e^{2\pi i b_3/N}, e^{-2\pi i b_3/N}\right)
\eeq
with
\beq
b_1=a_2+a_3, \quad b_2=a_3+a_1, \quad {\rm and} \quad b_3=a_1+a_2 \label{thetabos}
\eeq
 In general, $\gamma _{\theta,3}$ is an $(n+5) \times (n+5)$ matrix satisfying
 \beq
\gamma _{\theta,3}^N=\pm 1 \label{gam3N}
\eeq
In a Cartan-Weyl basis it can be written in the form
\beq
\gamma _{\theta , 3}= e^{-2\pi i V_3.H} \label{gamtheta3}
\eeq
with $H_I$ the Cartan generators of $U(n+5)$. Depending on the sign choice in (\ref{gam3N}), 
the vector $V_3$ has the form
\bea
 V_3&=& \frac{1}{N}(0^{n_0}, 1^{n_1}, ... , j^{n_j}, ...)  \label{V3p}\\
{\rm or } \ V_3 &=& \frac{1}{N}\left( \left(\frac{1}{2}\right)^{n_1},\left(\frac{3}{2}\right)^{n_3}, ... ,
 \left(j+\frac{1}{2}\right)^{n_{2j+1}}, ...\right)  \label{V3n}
\eea
where $ \sum _j n_j = n+5$ is the total number of D3-branes. 
Here we are using the shorthand  notation $j^{n_j}$ to denote $n_j$ entries $j$. 
In what follows we shall take $\gamma _{\theta,3}^N=+1$. However, a similar analysis 
can be made for the case $\gamma _{\theta,3}^N=-1$ with identical conclusions.

Gauge bosons arise from open strings that begin on a D3-brane and end on a D3-brane, i.e. they are (33)-sector states. 
The gauge bosons that survive the orbifold projection (i.e. are point-group invariant)
  are those associated with the (neutral) Cartan-Weyl generators, plus
   those associated with the 
charged generators  having root vectors $\rho _3 = (\underline{1,-1, 0^{n+3}})$ that  satisfy
\beq
\rho _3.V_3 =0 \bmod 1 \label{gauge}
\eeq
(The underlining signifies that all permutations of the underlined entries are to be included.)
Then  (\ref{V3p})  gives the surviving D3-brane gauge group as $\bigotimes _j U(n_j)$.
In particular,  we choose $V_3$ to have the form
\beq
V_3=\frac{1}{N}(x ^3, y ^2, d_1, d_2, ..., d_i,...) \label{SMgroup}
\eeq
where $x$ and $y$ are two different integers, and the single distinct entries $d_i$ also differ from $x$ and $y$, so that  
\beq
\forall i \ x \neq y \neq d_i \neq x \bmod N    \label{SMgroup2}
\eeq
Then $n_{x}=3$, $n_{y}=2$  and  $n_j=1 \ {\rm if} \ j \neq x,y$, and 
 the D3-brane gauge group is $U(3) \otimes U(2) \bigotimes _i U(1)_i$.

Point group invariance also implies that there is surviving matter on the D3-branes with root vectors $\rho _3$ of the above form 
that satisfy \cite{Aldazabal:2000sa, Alday:2002uc}
\bea
\rho _3.V_3&=&-\frac{a_{\alpha}}{N} \bmod 1 \quad {\rm (fermions)} \label{33fermions} \\
&=& \frac{b_r}{N} \bmod 1 \quad {\rm (scalars)}   \label{33scalars}
\eea
where $a_{\alpha} \ (\alpha = 1,2,3,4)$ are defined in (\ref{thetaferm}) and (\ref{thetaferm2}), and $b_r \ (r=1,2,3)$ 
are defined in (\ref{thetabos}). 
 This gives rise matter in bi-fundamental representations of the D3-brane gauge group:
 \bea
 \sum_{\alpha=1}^4 \sum _{j=0}^{N-1} ({\bf n}_j, \bar{\bf n}_{j+a_{\alpha}}) \quad {\rm (fermions)} \label{33fermspectrum}\\
 \sum_{r=1}^3 \sum _{j=0}^{N-1} ({\bf n}_j, \bar{\bf n}_{j-b_r}) \quad {\rm (scalars)} \label{33scalarspectrum}
 \eea
where all sub-indices are understood modulo $N$, and the fundamental (${\bf n}_j$) (anti-fundamental ($\bar{\bf n}_j$))
 representation of 
$SU(n_j)$ has respectively $+1(-1)$ units of the charge $Q_j$ associated with the $U(1)$ factor in $U(n_j)=U(1) \otimes SU(n_j)$.

 The obvious way to get three generations of quark doublets $Q_L$ each transforming as the ${\bf 3}$ of $SU(3)_c$ and the 
 ${\bf 2}$ of $SU(2)_L$ is to use the embedding (\ref{SMgroup}) of $\theta $ and to  choose
 \beq 
  (a_1,a_2,a_3, a_4) \equiv (a,a,a,c = -3a) \ (\bmod N) \label{aaa3a}
  \eeq
   Then, taking $y=x+a$, the $j=x$ contribution 
  to the sum (\ref{33fermspectrum})
  gives precisely the required three copies of the representation $({\bf 3}, \bar{\bf 2})$ of $SU(3)_c \otimes U(2)_L$. 
  Since these are the only ${\bf 3}$ representations in the standard model, we require also that 
  \beq
 \forall i \quad  x+a \neq d_i \neq x-3a  \ (\bmod N) 
 \eeq 
  The contributions to the sum (\ref{33fermspectrum}) from the terms with $j=x-a_{\alpha}$ give fermions transforming as 
  $\bar{\bf 3}$ representations of $SU(3)_c$:
\beq
3({\bf n} _{x-a}, \bar{\bf 3})+({\bf n} _{x+3a}, \bar{\bf 3}) \label{3bars}
\eeq
Provided that they have the correct weak hypercharges, these are potentially quark singlet states $u^c_L$ and $d^c_L$
and the corresponding states in other generations.
 However, to
 ensure the absence of unwanted $({\bf 2}, \bar{\bf 3}) $ states we require that $x-a \neq y \neq x+3a  \ (\bmod N)$, i.e. that
\beq
   2a \neq 0 \bmod N
\eeq
The states (\ref{3bars}) will appear in the fermionic spectrum as $(\bar{\bf 3}, {\bf 1})$ representations of $SU(3)_c \otimes SU(2)_L$ 
 if $x-a$ and/or  $x+3a$ is one of the entries $d_i$ in $V_3$ given in (\ref{SMgroup}). 
The number $n^{(33)}(\bar{\bf 3}, {\bf 1})$ of such states arising in the (33) sector is given by
\beq
n_{\rm fermions}^{(33)}(\bar{\bf 3}, {\bf 1})=\sum _i \left (3\delta _{d_i, x-a}+ \delta _{d_i, x+3a} \right)
\eeq
The $j=y$ and $j=y+b_r$ contributions to the  scalar spectrum (\ref{33scalarspectrum}) give states  transforming 
respectively as ${\bf 2}$ and $\bar{\bf 2}$ of $U(2)$.  With the values (\ref{aaa3a}) for the $a_{\alpha}$
we see from  (\ref{thetabos}) that 
\beq
(b_1,b_2,b_3)=(2a, 2a, 2a) \label{br2a}
\eeq
so the doublet states are
\beq
3({\bf 2 }, \bar{\bf n}_{y-2a})+3({\bf n} _{y+2a}, \bar{\bf 2}) \label{22bars}
\eeq
These are potentially Higgs doublets, and since $y=x+a$ we find that the numbers
 $n^{(33)}( {\bf 1}, {\bf 2})$ and $n^{(33)}( {\bf 1}, \bar{\bf 2}) $ of $( {\bf 1}, {\bf 2})$ and 
 $( {\bf 1}, \bar{\bf 2})$ representations of $SU(3)_c \otimes U(2)_L$  in the (33) sector  
 are
 \bea
 n_{\rm scalars}^{(33)}( {\bf 1}, {\bf 2})= 3\sum _i \delta _{d_i, x-a} \\
 n_{\rm scalars}^{(33)}( {\bf 1}, \bar{\bf 2}) = 3\sum _i \delta _{d_i, x+3a}
 \eea
 The weak hypercharge $Y$ is some {\it a priori} unknown linear combination of the $U(1)$ charges from 
 each factor in the gauge group:
 \beq
 Y= \alpha _x Q_x + \alpha _y Q_y + \sum _i \alpha_i Q_{d_i}+... \label{Y33}
 \eeq
 where the ... represents contributions from the D7-brane gauge groups, which will be discussed below.
 To get the correct weak hypercharges  for the 3 quark doublets $Q_L$, for the $u^c_L, d^c_L$ quark singlet states
  and the corresponding states 
 in other generations, as well as for the electroweak Higgs doublets that arise in the (33)-sector, we  take
 \beq
 Y=\frac{1}{6} Q_x +\frac{1}{2} \sum _{i \in I_d} Q_{d_i}-\frac{1}{2} \sum _{i \in I_u} Q_{d_i}+...
 \eeq
 where the sets $I_d$ and $I_u$ are non-overlapping and together include all of the $d_i$.
  We are dropping a contribution proportional to 
 $Q_x+Q_y+ \sum _i Q_{d_i}$ which is zero for all states.
 Clearly the (33) sector yields at least as many Higgs doublets as  quark  singlet $q^c_L$ states.

The (33)-sector fermion spectrum  generally makes the non-abelian gauge symmetries $SU(n_j)$ anomalous, and indeed 
in our model the colour-triplet (${\bf 3}$ and $\bar{\bf 3}$) fermions do make  $SU(3)_c$ anomalous.
This reflects the fact that a general collection of D3-branes has uncancelled RR tadpoles. The 
required twisted tadpole cancellation 
is effected locally by the introduction of D7$_r$-branes 
at the orbifold fixed point at which the D3-branes are located. The  D7$_r$-branes wrap the three large spatial dimensions 
and the four compact dimensions perpendicular to the $r$th complex plane, where $r=1,2,3$. 
As in (\ref{gamtheta3}), 
 the action of the $\mathbb{Z} _N$ point group on the  Chan-Paton 
 indices is encoded in a $u^r$-component vector of the form
 \bea
 V_{7_r} &=&\frac{1}{N}(0^{u^r_0}, 1^{u^r_1}, ... , j^{u^r_j}, ...)  \label{V7p} \\
  \hfill {\rm or} \ V_{7_r} &=& \frac{1}{N}\left( \left(\frac{1}{2}\right)^{u^r_1},\left(\frac{3}{2}\right)^{u^r_3}, ... ,
 \left(j+\frac{1}{2}\right)^{u^r_{2j+1}}, ...\right)  \label{V7n}
 \eea
 with $\sum _j u^r_j \equiv u^r$; (\ref{V7p}) (or (\ref{V7n})) applies when $\gamma _{\theta,7_r}^N=+1$ or (-1).
 The D7$_r$-brane gauge group is $\bigotimes _j U(u^r_j)$.  
 The introduction of D7$_r$-branes leads also to  (37$_r$)-  and (7$_r$3)-sector states which 
 arise from open strings 
 with one end on a D3-brane and the other on a D7$_r$-brane. 
 An advantage of the bottom-up models is that, since all of the matter is located at the (same) orbifold fixed point, 
 these models are free of the instanton-induced flavour changing neutral current problem that afflicts the 
 intersecting brane models.  
 In the (37$_r$)-  and (7$_r$3)-sectors   
   the point group generator is represented by the $(5+n+u^r)$-component  vector 
 \beq
 V_{37_r} \equiv (V_3;V_{7_r})
 \eeq
The (37$_r$)- and (7$_r$3)-states 
 are described by  weight vectors $\rho _{37_r}$ of the form
 \beq
 \rho _{37_r}= \left( \underline{\pm 1, 0^{n+4}}; \underline{\mp 1, 0^{u^r-1}} \right)
 \eeq
  In general the surviving  states satisfy \cite{Aldazabal:2000sa, Alday:2002uc}
  \bea
  \rho _{37_r}.V_{37_r}&=& -\frac{b_r}{2N} \ \bmod 1 \quad {\rm (fermions)} \\
  &=& \frac{b_s+b_t}{2N} \ \bmod 1   \quad r\neq s \neq t \neq r \quad {\rm (bosons)} 
  \eea
 For the three-generation model we are considering, the twists $b_r=2a$ given in (\ref{br2a}) are all even,
  in which case (\ref{V7p}) applies 
 when $\gamma _{\theta,3}^N=+1$, as we have assumed. The (37$_r$)+(7$_r$3)-sector spectrum is again given
  in terms of bi-fundamental representations, but now of the D3-brane and D7$_r$-brane gauge groups:
  \bea
  \sum _{j=0}^N \left[   ({\bf n}_j, \bar{\bf u}^r_{j+\frac{1}{2}b_{r}})
   +(\bar{\bf n}_j, {\bf u}^r_{j-\frac{1}{2}b_{r}})   \right]   \quad {\rm (fermions)}  \label{37fermions} \\
\sum _{j=0}^N \left[   ({\bf n}_j, \bar{\bf u}^r_{j-\frac{1}{2}(b_{s} +b_t)})
   +(\bar{\bf n}_j, {\bf u}^r_{j+\frac{1}{2}(b_{s}+b_t)})   \right]   \quad {\rm (scalars)}  \label{37scalars}
   \eea               
The contribution to (\ref{37fermions}) from the $j=x$ term gives fermions transforming as ${\bf 3}$ and $\bar{\bf 3}$
 representations  of $SU(3)_c$. 
To avoid unwanted ${\bf 3}$ states we must therefore require that $u^r_{x+a}=0$ for all $r=1,2,3$. The other term gives a 
number $n^{(37)}(\bar{\bf 3}, {\bf 1})$ of states transforming as the $  (\bar{\bf 3}, {\bf 1})$ 
representation of $SU(3)_c \otimes SU(2)_L$:
\beq
n_{\rm fermions}^{(37)}(\bar{\bf 3}, {\bf 1})= \sum _r u^r_{x-a}
\eeq
These too  are potentially 
 quark singlet states $u^c_L, d^c_L$ and the corresponding states in other generations, 
 provided that they have the correct weak hypercharges.
Similarly, the $j=y=x+a$ term in (\ref{37scalars})  gives  scalar states transforming 
 as $({\bf 1}, {\bf 2})$ and $({\bf 1}, \bar{\bf 2})$ representations of $SU(3)_c \otimes U(2)_L$:
 \bea
  n_{\rm scalars}^{(37)}( {\bf 1}, {\bf 2})= \sum _r u^r _{ x-a} \\
 n_{\rm scalars}^{(37)}( {\bf 1}, \bar{\bf 2}) = \sum _r u^r _{ x+3a}
 \eea
 and these are potentially electroweak Higgs doublets,  provided that they have the correct weak hypercharges.
 The correct hypercharges for both the quark singlet states and the electroweak Higgs doublets
  is achieved by including in (\ref{Y33}) the contributions from the $U(1)$ charges $Q_{j^r}$ of the $U(u^r_j)$ 
 factors in the D7-brane gauge group. We find that the (37)-sector states have the correct standard-model values if we take
  \beq
 Y=\frac{1}{6} Q_x +\frac{1}{2} \sum _{i \in I_d} Q_{d_i}-\frac{1}{2} \sum _{i \in I_u} Q_{d_i}
 +\frac{1}{2} \sum_{r,j^r \in J_d^r} Q_{j^r}-\frac{1}{2} \sum_{r,j^r \in J_u^r} Q_{j^r} \label{Y3337}
 \eeq
 where for each $r$ the sets $J_d^r$ and  $J_u^r$ are non-overlapping and include all values of $j^r$.
Thus, like the (33)-sector, the  (37)-sector too 
 yields at least as many Higgs doublets as there are  quark  singlet $q^c_L$ states. Consequently, the total number $n(q^c_L)$ 
 of quark singlet states is not greater than the number $n(H)+n(\bar{H})$ of Higgs doublets:
 \beq
 n(q^c_L) \leq n(H)+ n(\bar{H})
 \eeq 
 Since any standard-like model must have  three $u^c_L$ and three $d^c_L$ states,  so that $n(q^c_L)=6$, it follows that
 \beq
 n(H)+ n(\bar{H}) \geq 6
 \eeq
 and that the Higgs content cannot be that of the standard model, nor of its minimal supersymmetric extension, the MSSM.

In any case, it is well known that such orbifold models  cannot have a standard-like fermionic spectrum. 
Twisted tadpole cancellation 
implies the cancellation of the non-abelian anomalies \cite{Aldazabal:2000sa, Aldazabal:1999nu, Leigh:1998hj, Ibanez:1998qp}.
 Thus, after the inclusion of the  (37$_r$)+(7$_r$3)-sector
  matter, the same number of fermions in fundamental and anti-fundamental representations of 
 $SU(n_j)$ must be present for each  $SU(n_j)$. In other words the net charge $Q_j$ carried by fermions must vanish for every $j$. 
  However, having 3 copies of $Q_L$ transforming as the $({\bf 3}, \bar{\bf 2})$ representation of $SU(3)_c \otimes U(2)_L$
   generates 9 copies of the $\bar{\bf 2}$ representation of $U(2)_L$ having  a total $Q_y=-9$, so 
  to cancel it 
  the remaining fermions must include (at least) 9 copies of the ${\bf 2}$ representation of $U(2)_L$ having  a total $Q_y=9$. 
  Besides the  3 quark doublets, the standard model, of course, has precisely 3 copies of the ${\bf 2}$ representation of $SU(2)_L$,
   corresponding to the 3 lepton doublets $L$. Thus this method of generating the quark doublets inevitably entails the 
   existence of  6 vector-like lepton doublets $L+\bar{L}$, not present in the standard model, and  the first 
   bottom-up attempts \cite{Aldazabal:2000sa, Bailin:2000kd}
    to get the standard model indeed suffered from this defect.

   
The only escape is to arrange that three quark doublets do not all have the same $Q_y$ charge \cite{Ibanez:2001nd}.
We require $2({\bf 3}, \bar{\bf 2}) +({\bf 3}, {\bf 2})$ 
representations of $SU(3)_c \otimes U(2)_L$ (or equivalently   $({\bf 3}, \bar{\bf 2}) +2({\bf 3}, {\bf 2})$) 
which have a total $Q_y=-3$ (or $Q_y=+3$). This can be cancelled by 3 lepton doublets transforming respectively as the 
 ${\bf 2}$ (or $\bar{\bf 2}$) representation of $U(2)_L$. Getting both representations is possible 
 only when the $\mathbb{Z} _N$ orbifold singularity at which the D3-branes are situated is on an orientifold plane 
 \cite{Ibanez:2001nd, Alday:2002uc}. In the orientifolds that we are considering the point group 
   quotienting  the torus $T^6$ is  enlarged. It is generated by $\{\theta, \Omega \}$ where, as before, $\theta$ is the 
   generator of $\mathbb{Z} _N$ and the new generator $\Omega$ is the world-sheet parity operator. 
   Thus the complete orientifold group is $\mathbb{Z} _N + \Omega \mathbb{Z} _N$. The requirement of invariance 
   under the action of  the extra generator $\Omega$ restricts the form  of the embedding $\gamma _{\theta ,3}$ 
   given in (\ref{gamtheta3}), and hence the form of $V_3$ given in (\ref{V3p},\ref{V3n}). $\Omega$ acts on the
   diagonal matrix $\gamma _{\theta ,3}$ as complex conjugation, and (for the case that 
   $\gamma _{\theta ,3}^N=+1$) the invariance requires \cite{Aldazabal:1998mr}
    that $V_3$ has the form 
   \beq
   V_3=(\tilde{V_3};-\tilde{V_3})
   \eeq
   and $ \tilde{V_3}$ is given by
   \beq
   \tilde{V_3}=\frac{1}{N}(0^{n_0}, 1^{n_1},..., P^{n_P})
   \eeq
   with $P\equiv \left[\frac{N}{2}\right]$ the largest integer not greater than $\frac{N}{2}$. Then the 
   gauge group factor $U(n_j)$ is the same as $U(n_{-j}) \equiv U(n_{N-j})$ and the two are exchanged under the action of $\Omega$. 
   Invariance requires that they are identified. The (33)-sector spectrum may be calculated as before, using equations (\ref{gauge}, 
   \ref{33fermions}, \ref{33scalars}), but with $V_3$ replaced by $\tilde{V_3}$ and now with 
   \beq
   \rho _3=(\underline{ \pm1, \pm1, 0,0, ...}) \label{rho3fold}
   \eeq
   and all four combinations of signs allowed. For a general embedding the D3-brane gauge group is 
   $SO(2n_0) \bigotimes _{j =1}^P U(n_j)$. The extra states included in (\ref{rho3fold}) lead to extra 
    fermions and scalars  surviving the projection. Instead of (\ref{33fermspectrum}) and (\ref{33scalarspectrum}), we 
    now have
  \bea
 \sum_{\alpha=1}^4 \sum _{j=0}^{N-1} \left[ ({\bf n}_j, \bar{\bf n}_{j+a_{\alpha}}) +({\bf n}_j, {\bf n}_{-j-a_{\alpha}}) 
 +(\bar{\bf n}_j, \bar{\bf n}_{-j+a_{\alpha}}) + \right. \nonumber \\
\left. +({\bf n}_j \times {\bf n}_j )_a \delta _{2j, -a _{\alpha}}+
 (\bar{\bf n}_j \times \bar{\bf n}_j )_a \delta _{2j, a _{\alpha}} 
 \right] \quad {\rm (fermions)} \label{33fermspectrum2} \\
 \sum_{r=1}^3 \sum _{j=0}^{N-1} \left[ ({\bf n}_j, \bar{\bf n}_{j-b_r}) 
  +({\bf n}_j, {\bf n}_{-j+b_r}) +(\bar{\bf n}_j, \bar{\bf n}_{-j-b_r}) + \right. \nonumber \\
 \left. +({\bf n}_j \times {\bf n}_j )_a \delta _{2j, b_r}+ 
 (\bar{\bf n}_j \times \bar{\bf n}_j )_a \delta _{2j, -b_r} \right] 
   \quad {\rm (scalars)} \label{33scalarspectrum2}
 \eea   
Here $({\bf n}_j \times {\bf n}_j )_a$ is the antisymmetric  $\frac{1}{2}n_j(n_j-1)$-dimensional representation of $U(n_j)$ 
that arises in the product ${\bf n}_j \times {\bf n}_j $, and similarly for   $(\bar{\bf n}_j \times \bar{\bf n}_j )_a $.

When $\tilde{V_3}$ has the form given in  (\ref{SMgroup}, \ref{SMgroup2}), namely
\beq
\tilde{V}_3=\frac{1}{N}(x ^3, y ^2, d_1, d_2, ..., d_i,...) \label{SMgroup3}
\eeq
with $x$ and $y$ non-zero and
\beq
x \neq y \neq d_i \neq x \bmod N  \ (\forall i) \label{SMgroup4}
\eeq
then, as before,   the gauge group is $U(3) \otimes U(2) \bigotimes _i U(1)_i$.
    Also as before, quark doublets  $Q_L$ transforming as  the $({\bf 3}, \bar{\bf 2})$ representation of 
    $SU(3)_c \otimes U(2)_L$  can  only arise from
     the $j=x$ contribution to the first term of (\ref{33fermspectrum2}), and to get 
     2 copies two of the $a_{\alpha}$ must be equal. So we take
\bea
(a_1, a_2, a_3, a_4) &=& \left(a, a, b, c=-(2a+b)\right)  \ (\bmod N) \label{aalpha3} \\
(b_1,b_2,b_3) &=& (a+b, a+b, 2a) \label{br2}  \ (\bmod N)
\eea
with
\beq
a \neq b \neq c \neq a \ (\bmod N) \label{abcneq}
\eeq 
so as to avoid the occurrence of three identical quark doublets.
Then,  the required 2 copies of $Q_L$ arise if $y=x+a$. The  
 $({\bf 3}, {\bf 2})$ fermion representation of  $SU(3)_c \otimes U(2)_L$
arises from the $j=x$ contribution to the second term of (\ref{33fermspectrum2}), and we get
the required single copy if $y=-x-b$. 
The definition (\ref{Y3337}) of the weak hypercharge again ensures that all of these states have the 
correct standard-model weak hypercharge $Y=\frac{1}{6}$. 
 (We also get the required $({\bf 3}, {\bf 2})$ fermion representation if  $y=-x-c$, 
 but the physics in the two cases is, of course, identical.) 
 Note that in the orientifold case the contribution to $Y$ proportional to 
$Q_x+Q_y+ \sum _i Q_{d_i}$ is no longer automatically zero for all states.
 However, since we require that both the $({\bf 3}, \bar{\bf 2})$ 
 and the  $({\bf 3}, {\bf 2})$ fermion states have the same weak hypercharge, this contribution must be absent.
  Thus $\tilde{V_3}$ has the form:
\beq
\tilde{V_3} =\frac{1}{N} \left[ \left( -\frac{a+b}{2} \right)^3, \left( \frac{a-b}{2} \right)^2, d_1,d_2,... \right] \label{Vtild3}
\eeq
with all of the entries non-zero and unequal in order to avoid enlargement of the gauge group. 
For the same reason we also require that
\bea
-\frac{1}{2}(a+b) &\neq & +\frac{1}{2}(a+b), \ \pm \frac{1}{2}(a-b), \  \pm d_i  \quad (\bmod N) \\
-\frac{1}{2}(a-b) &\neq & +\frac{1}{2}(a-b), \ \pm d_i \quad (\bmod N) \\
-d_i &\neq & d_i,  \ \pm d_j \ (j\neq i) \quad (\bmod N) \label{V3ineq} 
\eea
To ensure that there is no unwanted 
vector-like quark doublet matter, we require further that
\beq
2a \neq 0, \quad 2b \neq 0 \ (\bmod N)
\eeq 
Since we are assuming that $\gamma _{\theta ,3}^N=+1$, as before, then $a-b=0 \bmod 2=a+b$. 
Quark singlet states $\bar{q}^c_L$ transforming as the $({\bf 3}, {\bf 1})$ representation
 and  $q^c_L$ states transforming as the $(\bar{\bf 3}, {\bf 1})$ 
representation of  $SU(3)_c \otimes SU(2)_L$  can  also arise from the $j=x$ and $j=x-a_{\alpha}$ 
contributions to (\ref{33fermspectrum2}). 
Proceeding as before, we find that the number of such states is given by
\bea
n^{(33)}_{\rm fermions}({\bf 3}, {\bf 1})&= &2\delta _{2a+b,0}+\delta _{a+2b,0}+ \sum _i \left[ \delta _{d_i, \frac{1}{2}(5a+3b)} 
+ \delta _{d_i, -\frac{1}{2}(5a+3b)} \right] \\
n^{(33)}_{\rm fermions}(\bar{\bf 3}, {\bf 1})&=&\delta _{3a+2b,0}+ 3\sum _i  (\delta _{d_i, \frac{1}{2}(3a+b)} 
+ \delta _{d_i, -\frac{1}{2}(3a+b)}) + \nonumber \\
& \hfill & +\sum _i (\delta _{d_i, \frac{1}{2}(a+3b)}
  + \delta _{d_i, -\frac{1}{2}(a+3b)} )  \label{n333bar}
\eea
Again, (\ref{Y3337})  ensures that these states have the 
correct standard-model weak hypercharge assignments. To exclude the unwanted $\bar{q}^c_L$ states we must take
\beq
2a+b \neq 0 \neq a+2b  \ {\rm and} \  \forall i \ \pm d_i \neq \frac{1}{2}(5a+3b) \  (\bmod N)
\eeq
The embedding (\ref{aalpha3}), and the spectrum, is supersymmetric in the special case that $c=0=2a+b \bmod N$.
Thus the first of these inequalities excludes the supersymmetric embedding, 
and hence the possibility of obtaining the MSSM \cite{Alday:2002uc}. 

The contributions to  the scalar spectrum (\ref{33scalarspectrum2}) from $j=y$ and $ j=y+b_r$ 
give potential Higgs doublets with the correct hypercharges. We find that 
\bea
n^{(33)}_{\rm scalars}({\bf 1}, {\bf 2})=\sum _i \left[ 2(\delta _{d_i, \frac{1}{2}(a+3b)} +\delta _{d_i, -\frac{1}{2}(a+3b)}) 
+ \delta _{d_i, \frac{1}{2}(3a+b)}+\delta _{d_i, -\frac{1}{2}(a+3b)} \right]  \label{n332}\\
n^{(33)}_{\rm scalars}({\bf 1}, \bar{\bf 2})=\sum _i \left[ 2(\delta _{d_i, \frac{1}{2}(3a+b)} +\delta _{d_i, -\frac{1}{2}(3a+b)}) 
+ \delta _{d_i, \frac{1}{2}(5a-b)}+\delta _{d_i, -\frac{1}{2}(5a-b)} \right] \label{n332bar}
\eea
Comparing (\ref{n332}) and (\ref{n332bar}) with (\ref{n333bar}), we see that 
\beq
n^{(33)}_{\rm scalars}({\bf 1}, {\bf 2})+n^{(33)}_{\rm scalars}({\bf 1}, \bar{\bf 2}) \geq n^{(33)}_{\rm fermions}(\bar{\bf 3}, {\bf 1}) 
-\delta _{3a+2b,0} \label{n33HHbarqcL}
\eeq

The (37$_r$) + (7$_r$3)-sector spectrum is calculated in a similar way, and leads to
\bea
 \sum _{j=0}^N \left[   ({\bf n}_j, \bar{\bf u}^r_{j+\frac{1}{2}b_{r}}) + ({\bf n}_j, {\bf u}^r_{-j-\frac{1}{2}b_{r}})
  \right. & +& \left. (\bar{\bf n}_j, {\bf u}^r_{j-\frac{1}{2}b_{r}})  + (\bar{\bf n}_j, \bar{\bf u}^r_{-j+\frac{1}{2}b_{r}})
    \right]   \ {\rm (fermions)}  \label{37fermions2} \\
\sum _{j=0}^N  \left[   ({\bf n}_j, \bar{\bf u}^r_{j-\frac{1}{2}(b_{s} +b_t)}) \right. & +& \left.  ({\bf n}_j, {\bf u}^r_{-j+\frac{1}{2}(b_{s} +b_t)})
   +(\bar{\bf n}_j, {\bf u}^r_{j+\frac{1}{2}(b_{s}+b_t)}) \right. + \hfill \nonumber \\
    \hfill & +& \left. \hfill (\bar{\bf n}_j, \bar{\bf u}^r_{-j-\frac{1}{2}(b_{s}+b_t)})
    \right]   \ {\rm (scalars)}  \label{37scalars2}
 \eea          
Then
\bea
n^{(37)}_{\rm fermions}({\bf 3}, {\bf 1})&=& \sum _{r=1,2} 2u^r_0+ u^3_{\frac{1}{2}(a-b)}+ u^3_{-\frac{1}{2}(a-b)} \\
n^{(37)}_{\rm fermions}(\bar{\bf 3}, {\bf 1})&=& \sum _{r=1,2} (u_{a+b}^r+ u_{-(a+b)}^r)
+ u^3_{\frac{1}{2}(3a+b)}+ u^3_{-\frac{1}{2}(3a+b)} \label{n373bar}\\
n^{(37)}_{\rm scalars}({\bf 1}, {\bf 2})&=& \sum _{r=1,2} (u_{a+b}^r+ u_{-(a+b)}^r) 
+ u^3_{\frac{1}{2}(a+3b)}+ u^3_{-\frac{1}{2}(a+3b)} \label{n372}\\
n^{(37)}_{\rm scalars}({\bf 1}, \bar{\bf 2})&=& \sum _{r=1,2} (u_{2a}^r+
 u_{-(2a)}^r) + u^3_{\frac{1}{2}(3a+b)}+ u^3_{-\frac{1}{2}(3a+b)} \label{n372bar}
\eea
Clearly we require that
\beq
u^1_0=u^2_0=u^3_{\frac{1}{2}(a-b)}= u^3_{-\frac{1}{2}(a-b)}=0
\eeq
in order to exclude exclude unwanted $\bar{q}^c_L$ states. 
The definition (\ref{Y3337})  again ensures that these states have the 
correct standard-model weak hypercharge assignments.
Comparing (\ref{n372}) and (\ref{n372bar}) with (\ref{n373bar}), we see that
\beq
n^{(37)}_{\rm scalars}({\bf 1}, {\bf 2})+n^{(37)}_{\rm scalars}({\bf 1}, \bar{\bf 2}) \geq n^{(37)}_{\rm fermions}(\bar{\bf 3}, {\bf 1}) 
\eeq
Combining this with the analogous (33)-sector result (\ref{n33HHbarqcL}), we conclude that 
the total number $n(H)+n(\bar{H})$ of Higgs doublets and the total number $n(q^c_L)$ of quark singlets satisfy
\beq
n(H)+n(\bar{H}) \geq n(q^c_L)-\delta _{3a+2b,0}
\eeq
As before, any standard-like model requires that  there are $n(q^c_L)=6$ quark singlet states.
Thus again the Higgs content cannot be that of the standard model, nor that of the MSSM.

We have shown that non-minimal Higgs content is unavoidable in  models 
which seek to replicate the standard model spectrum, or that of the MSSM, 
by starting with D3-branes situated at a singular point of a $\mathbb{Z}_N$ orbifold or orientifold. This conclusion is 
independent of the order $N$ of the point group $\mathbb{Z}_N$. It is also independent of the
 gauge groups living on the D7-branes that have to be introduced to complete the spectrum, and which are 
 in any case inescapable to ensure cancellation of the RR tadpoles. 
 Of course, tadpole cancellation, or equivalently cancellation of the non-abelian anomalies,  constrains 
 these groups, and it 
may be that  even stronger lower limits on the Higgs content can be obtained if these constraints are imposed.
 However, we have not explored this point further. As already noted, the orbifold models are  known to be 
deficient in other respects, in particular their necessity for vector-like lepton-doublet matter. Orientifold 
models do not have this affliction, but to avoid vector-like quark-singlet matter we must take a non-supersymmetric 
embedding of the point group. Even so, vector-like ``squark''-singlet matter {\em is} unavoidable. (We are 
using the term ``squark'' loosely here since there is no supersymmetry; we mean simply scalars tranforming 
as the $({\bf 3},{\bf 1})$ representation of $SU(3)_c \otimes SU(2)_L$.)
 The $j=x$ contribution to the last term   of the scalar spectrum (\ref{33scalarspectrum2}) includes 2 copies of 
$(\bar{\bf 3} \times \bar{\bf 3})_a={\bf 3}$, since for this term $2j=-(a+b)=-b_1=-b_2$. However, 
we regard the appearance of  vector-like squarks as a less serious defect than the appearance of vector-like fermions, because 
large masses for the former can be generated by strong radiative corrections \cite{Cremades:2002dh}, whereas fermion 
masses are protected by chiralty considerations.
There is a further objection to   the orbifold and orientifold models of the type that we are considering, which is that 
the (37)+(73)-sectors contains matter that transforms non-trivially with respect to both the standard model gauge group 
and the D7-brane gauge groups. Since there is no evidence for any  gauge symmetry other than that of the standard model 
gauge group,  if these models were to occur in nature, then it must be that the non-abelian D7-brane gauge groups
 are completely broken.  
At most a single surviving gauged $U(1)$ can survive, with other $U(1)$s surviving only as global symmetries after taking account of
Green-Schwarz terms and the possibility of scalars in the (77)-sectors acquiring non-zero vacuum expectation values (VEVs) to remove 
unwanted $U(1)$ gauge groups.
 In principle, by a judicious choice of Wilson lines, it might be possible 
to arrange this. However, such symmetry breaking does not change the numbers $u^r_j$ of  standard-model representations 
that occur in the (37$_r$)+(7$_r$3)-sectors, and so will not change the numbers of quark singlets or Higgs doublets. 
 This is because the D3-branes are situated at the origin, and the massless (37$_r$)+(7$_r$3)-sector states therefore have both
  ends of the string at the origin. Thus, the influence of the Wilson lines is not felt. 
  We have already observed that this same feature also ensures that the problem with instanton-induced
   flavour changing neutral currents that afflicts 
  the intersecting brane models does not arise in these models. However, it might also make the emergence of a realistic 
  mass hierarchy difficult to achieve.   
As observed in the introduction, another objection to  non-supersymmetric models is that cancellation of RR tadpoles does not guarantee 
cancellation of the NSNS tadpoles. Thus although the complex structure moduli are fixed by the point group symmetry 
in any (orbifold or) orientifold model, 
the stabilsation of the dilaton remains problematic, as in the (non-supersymmetric) intersecting brane models. 
In any case, the conclusion that {\em all} such models have non-minimal Higgs content means 
that after electroweak symmetry breaking they are all afflicted with tree-level flavour-changing neutral currents
 mediated by Higgs exchange. 
The  severity of the experimental limits  on these processes means that these models too are effectively dead. The only 
escape from this conclusion that we can see is if the (77)-sector VEVs are large and effectively remove some of the Higgs doublets 
from the low-energy spectrum. Such VEVs cannot affect the number $n(q^c_L)$ of quark singlet states, because mass terms for them 
can only arise from VEVs for the electroweak Higgses in the (33)- and (37)-sectors. However, in the absence of supersymmetry, 
the calculation of the required VEVs entails the calculation of the $\phi ^4$ terms in the effective potential for the 
(77)-sector scalars, and we have not attempted this. Even if it is possible in principle, 
it seems unlikely that the survival of a single $H+ \bar{H}$ pair would be generic.

A similar analysis can be made for the left-right symmetric and Pati-Salam models with gauge groups 
$SU(3)_c \otimes SU(2)_L \otimes SU(2)_R$ and $SU(4)_c \otimes SU(2)_L \otimes SU(2)_R$ respectively. 
In the latter case, lepton number is the fourth ``colour'' and  a single fermionic generation (including a right-chiral neutrino)
 is contained in the representations 
$({\bf 4}, {\bf 2}, {\bf 1}) + (\bar{\bf 4}, {\bf 1}, {\bf 2}) $. 
In the former  ${\bf 4} \in SU(4)_c$ is ${\bf 3} + {\bf 1} \in SU(3)_c$ and $\bar{\bf 4}=\bar{\bf 3} + {\bf 1}$. 
In both cases  the Higgs bosons are required to be in the $({\bf 1}, {\bf 2}, {\bf 2}) \in SU(n_c) \otimes SU(2)_L \otimes SU(2)_R$ 
where $n_c = 3 \ {\rm or} \ 4$ is the number of colours.
To get the required gauge group, instead of (\ref{SMgroup}) we take 
\beq
V_3=\frac{1}{N}(x ^{n_c}, y ^2,z^2, d_1, d_2, ..., d_i,...) \label{nc22group}
\eeq
where $x,y,z$ and $d_i$ are all different integers ($\bmod N$). In the orbifold case we take the $a_{\alpha}$ 
as in (\ref{aaa3a}). Then choosing $y=x+a$, as before, gives the required three copies of $({\bf n_c}, \bar{\bf 2}, {\bf 1}) \in 
U(n_c) \otimes U(2)_L \otimes U(2)_R$. The right-chiral states transforming as  $(\bar{\bf n} _c, {\bf 1}, {\bf 2})$ 
must all be in the 33 sector, and we get  the required three copies when $z=x-a=y-2a$. Then, unavoidably, there is non-minimal 
Higgs content since there are  
three copies of  $({\bf 1}, {\bf 2}, \bar{\bf 2})$ in the bosonic sector.    In the orientifold case, we take $a_{\alpha}$ 
as in (\ref{aalpha3}). Taking $\tilde{V}_3$ to have the form (\ref{nc22group}), where $x=-\frac{1}{2}(a+b)$ and 
$y=\frac{1}{2}(a-b)$ as in (\ref{Vtild3}), then gives the required $2({\bf n_c}, \bar{\bf 2}, {\bf 1})+({\bf n_c}, {\bf 2}, {\bf 1})$ 
left-chiral matter content.  The only way to get the required three copies of 
$(\bar{\bf n}_c, {\bf 1},{\bf 2}) \in SU(n_c)\otimes SU(2)_L \otimes SU(2)_R$ then is if $z=\pm \frac{1}{2}(3a+b)$; the 
positive sign gives $2(\bar{\bf n} _c, {\bf 1}, \bar{\bf 2})+(\bar{\bf n} _c, {\bf 1}, {\bf 2})
 \in U(n_c) \otimes U(2)_L \otimes U(2)_R$, while the negative sign gives the conjugate $U(2)_R$ representations. 
 As before, this fixes the Higgs boson content. 
 In the former case we get $2({\bf 1}, \bar{\bf 2}, {\bf 2})+({\bf 1}, {\bf 2}, {\bf 2})$, while  in the latter 
 the conjugate $U(2)_R$ representations arise. Either way the Higgs content is again non-minimal, since we have more than one
 copy of one of the multiplets. It is also insufficient to give  Yukawa couplings for all left- and right-chiral states.

\section*{Acknowledgements} 
We are grateful to Gabriele Honecker for helpful correspondence. This research is supported in part by PPARC.

\end{document}